\documentclass[a4paper]{article}
\usepackage[english]{babel}
\usepackage[utf8]{inputenc}
\usepackage{amsmath}
\usepackage{graphicx}
\usepackage[colorinlistoftodos]{todonotes}
\usepackage{hyperref}

\usepackage{amsfonts}       
\usepackage{nicefrac}       
\usepackage{microtype}      
\usepackage{lipsum}
\usepackage{graphicx}
\graphicspath{ {./images/} }

\usepackage{graphicx}
\usepackage{subcaption}
\usepackage{amssymb}
\usepackage{amsmath}

\title{Conformable 1D steady-state Navier-Stokes equations to describe flow through porous media}
\author{Santos-Moreno, M., Fernández-Anaya G., 
Valencia-Negrete C.V.}
\date{}
\begin{document}
\maketitle

\begin{abstract}
From the definition of a generalized conformable spatial derivative, an exponential conformable function with three parameters $(a,b,\alpha)$ is proposed for a viscous and an inertial-viscous steady-state Navier-Stokes 1D models, obtaining analytical solutions for both generalized conformable models. The conformable models' parameters are optimized to compare the viscous model to a Darcian 1D flow and the inertial-viscous model to a non-Darcian 1D model for a specific range of Darcy numbers  $(1\times10^{-2}<Da\leq 1)$. Velocity profiles for the porous medium and the conformable model are computed and compared, showing that the generalized conformable Navier Stokes 1D models describe the flow through a porous medium, for both Darcian and non-Darcian flow, without including a Darcy term or macroscopic porous characteristics. 
\end{abstract}


\section{Introduction}
\label{S:I}
Fractional calculus has proven to be a successful tool to model phenomena owning long-range memory special effects that traditional integer-order calculus finds difficult to describe: characterization of anomalous diffusion phenomena, constant-order fractional diffusion equations, viscoelasticity, among other areas \cite{AtanganaBook}. Although the fractional derivatives display several advantages, they do not satisfy all the operational behaviors than the classical first derivative, such as product rule, quotient rule, chain rule, and semigroup property \cite{atangana15}. These lead to the development of the local fractional derivative, which properties coincide with classical integer derivative. Khalil \cite{KHALIL14} introduced a new definition for a local fractional derivative called conformable derivative, which portrays the most characteristics of classical derivative \cite{ZHAO}, \cite{RN207}. Because of its mathematical properties, conformable derivative has been used in various field such as anomalous diffusion \cite{AnomDiffConfor},fluid mechanics \cite{NSCtim}, \cite{Burgersfluid}, among several others \cite{ALHARBI201918}, \cite{yavuz2018conformable}, \cite{yepez2018first}, \cite{hyder2020exact}. The conformable derivative interest has been increasing, and it is worthwhile to explore its range for different physical situations. Zhao et al. \cite{ZHAO} analyzed some physical interpretations of conformable derivatives in terms of a generalized conformable function for spatial variable derivative. However, the majority of the studies about conformable derivative are still focused on time variable solutions, mathematical methods, and numerical solutions \cite{AKBULUT2018876}, \cite{anderson2015newly}, \cite{hosseini2017new}, \cite{benkhettou2016conformable}. In transport phenomena, the conformable derivative approach has been applied to the dispersion diffusion equation, generally focused on pollutant mass transfer with time-conformable derivative \cite{CHAUDHARY20}, \cite{YANG2020106330}, \cite{Chen}, \cite{Yang2019AnalyticalSO}.

Fluid flow through porous media is an important research area due to its multiple applications. Some of the recent investigations have been focused on simulations and solving methods for multiphase flows \cite{CLARKE20}, \cite{khoei2021}, \cite{maruvsic20} and interactions between flowing fluid and permeable solids \cite{sherwood20}, \cite{KANAUN2020}, most of them describing viscous and inertial phenomena. The equations describing flow through a porous medium can be classified as Darcian and non-Darcian models. Darcian models describe the flow under the influence of viscous terms (diffusive process with negligible convection), while non-Darcian models also consider the influence of the convective term \cite{HADMAN91}. Darcy's Law and Brinkman's model represent the Darcian flow, while  Darcy-Lapwood-Brinkman (DLB), Darcy-Forchheimer (DF), Darcy-Forchheimer-Brinkman (DFB) are members of non-Darcian models. The models has been widely applied and its range of validity has been thoroughly discussed \cite{HADMAN91}, \cite{HAMDAN1994203}, \cite{Neale}, \cite{Merabet}, \cite{sherwood20}. Conformable calculus has also applied to describe fluid flow through a porous medium, focusing on the time variable derivative for non Darcian flows or precisely mathematical proofs \cite{Lei}, \cite{IYIOLA2017}, \cite{li2020existence}, \cite{yang2018}. The purpose of this work is to introduce a generalized form of a spatial-variable conformable derivative to two steady-state 1D Navier-Stokes equations, analyzing its effect on the velocity profiles and comparing it to the 1D flow through a non-homogeneous space.The viscous conformable model is compared to a Brinkman 1D model, and a 1D DLB model is selected to compare with the inertial-viscous conformable model. A good agreement with the conformable solution corresponding to the classical porous medium model is found.

This work is organized as follows. In Section \ref{S:Pre}, the general definition of a generalized conformable derivative and its basic properties are introduced. Section \ref{S:NewPo} presents the analytical solutions for viscous flow and an inertial-viscous flow in the steady-state 1D Navier Stokes models and their comparison to porous medium classical solutions. The conformable derivative is introduced to Navier Stokes steady-state 1D models and compared to the porous medium models in Section \ref{S:GCD_Model}. Finally, some conclusions are presented in Section \ref{S:Concl}.

\section{Preliminaries}
\label{S:Pre}
Zhao et al \cite{ZHAO} introduced the Generalized Conformable Fractional Derivative (GFCD), as an extension of the classical space derivative and the conformable fractional derivative (CFD) for a function $f(u):[0,\infty)\to \mathbb{R}$, for all $u>0$ and $\alpha \in (0,1]$
\begin{equation}
\label{eq:comf}
D^\alpha_\psi f(u)= \lim_{\epsilon\to 0} \frac{f(u+\epsilon \psi (u,\alpha))-f(u)}{\epsilon}
\end{equation}
where $\psi(u,\alpha)$ is the fractional conformable generalized function, a continuous real function, depending on $u$ and satisfying:
\begin{center}
   $\psi(u,1)=1$,\\ 
   $\psi (u,\alpha) \neq \psi(u,\beta) \ \alpha \neq\beta \in (0,1]$
\end{center}
When $\psi(u,\alpha)=1$, $D^\alpha_\psi f(u)$ degenerates to the usual first-order derivative and has no relationship with fractional order $\alpha$.
Theorems are proved for this generalized conformable derivative, including \cite{ZHAO}:
\begin{align*}
(1)&\ D^\alpha_\psi (af+bg)= (a D^\alpha_\psi f+ bD^\alpha_\psi g), \text{for all}\ a,b \in \mathbb{R}.\\
(2)&\ D^\alpha_\psi(fg)=fD^\alpha_\psi g+ g D^\alpha_\psi f.\\
(3)&\ D^\alpha_\psi \left(\frac{f}{g}\right)=\frac{g D^\alpha_\psi f- f D^\alpha_\psi g}{g^2}.\\
(4)&\ \text{Chain rule\ } D^\alpha_\psi (f\circ g)=f'(g(u)) D^\alpha_\psi g(u).\\
(5)&\ \text{If $f$ is differentiable, then}\ D^\alpha_\psi (f)=\psi (u,\alpha)\frac{df}{du}.
\end{align*}
For more basic properties and main results on conformable derivatives, refer to \cite{ZHAO}, \cite{RN207}, \cite{KHALIL14}.
Some exponential, logarithmic, polynomial, trigonometric functions, among others, can satisfy these characteristics. Some examples \cite{ZHAO} of the conformable function $\psi(u,\alpha)$ are listed as follows:
\begin{center}
    linear: $\psi(u,\alpha) = \alpha k+ b$, where k, b are constant numbers, \\ 
    exponential: $\psi (u,\alpha)= a^{(1-\alpha)h(\alpha)}$, where $h(p)$ is a polynomial function.
\end{center}
\begin{figure}[htb!]
    \centering
    \begin{subfigure}[b]{0.47\textwidth}
    \includegraphics[width=\linewidth]{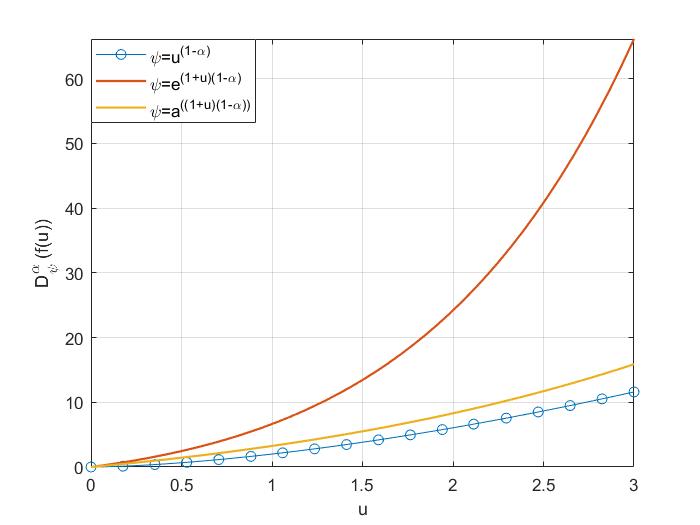}
    \caption{\small Function $u^2$}
    \label{fig:u2}
    \end{subfigure}\hfill
\begin{subfigure}[b]{0.47\textwidth}
    \includegraphics[width=\linewidth]{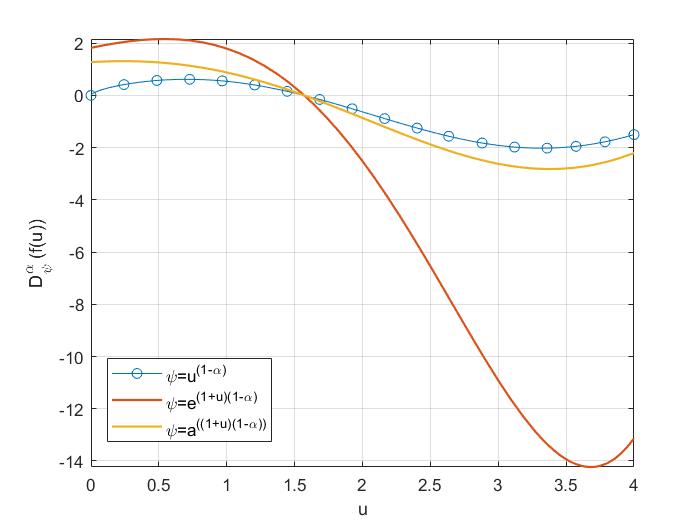}
    \caption{\small Function $sin(u)$}
    \label{fig:sinu}
    \end{subfigure}
    \caption{Examples of Generalized Conformable Derivative }
    \label{fig:fcn}
    \end{figure}
To show the influence of conformable functions in the GCFD of a defined function, two examples are shown in fig. (\ref{fig:fcn}), for a polynomial function $f_1(u)=u^2$ and a trigonometric function $f_2(u)=sin(u)$. Conformable functions selected were Khalil definition (blue marker) and two different types of exponential functions, with a fixed fractional value of $\alpha=0.4$. The selection of the conformable function in the GCFD will be based on how the derivative modifies the differential equation's behavior representing the flow to solve. \\
\section{Parallel and fully-developed flow through a straight channel}
\label{S:NewPo}
The fundamental governing equations for Newtonian fluid flow are the Navier-Stokes equations (N-S), representing the conservation of momentum and describing the flow through a free space.A steady-state 1D viscous and a steady-state 1D inertial-viscous N-S models are compared to a porous medium models to observe the effect of the porous matrix on the velocity profiles.To describe the flow of an incompressible fluid through porous media, the averaging theorem of the momentum equations, by Slattery-Withaker, has been successfully applied for some single-phase porous medium from a macroscopic perspective \cite{HAMDAN1994203}, \cite{Slattery}, \cite{WHITAKER1966291}. A 1D Brinkman model and a 1D DLB have been selected to compare with the viscous and the inertial-viscous flows, respectively. Afterward, a comparison between the conformable model and the porous medium is proposed.
\begin{figure}[htb!]
\begin{subfigure}[b]{0.45\textwidth}
\centering
\includegraphics[width=\linewidth]{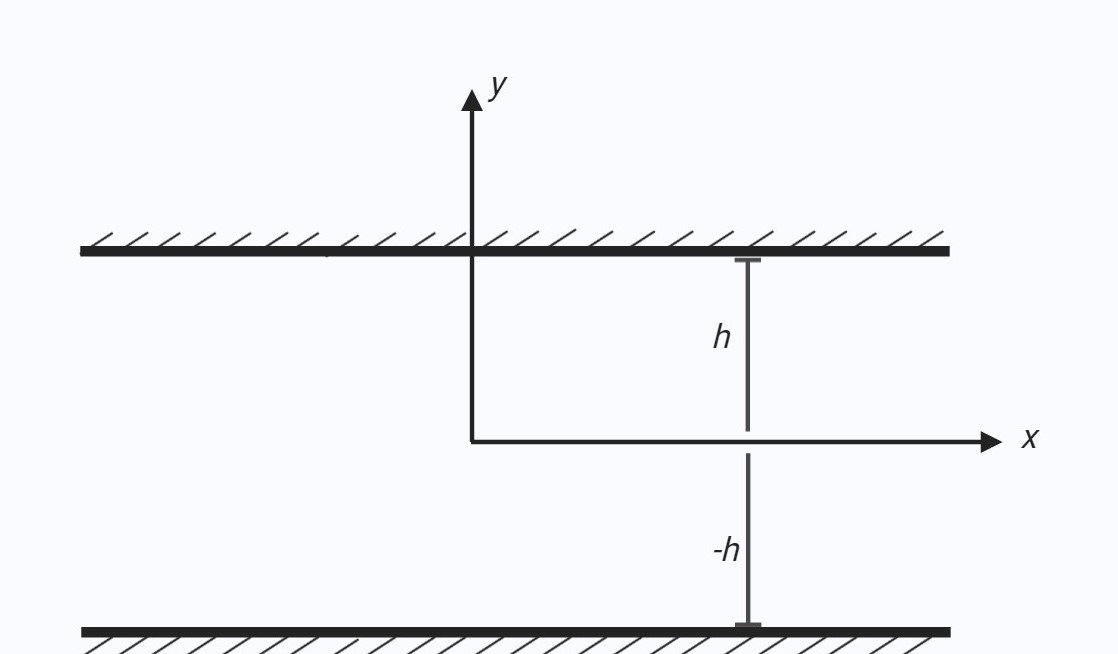}
\label{subfig:freespace}
\caption{For Navier-Stokes 1D equation}
\end{subfigure}\hfill
\begin{subfigure}[b]{0.45\textwidth}
\centering
\includegraphics[width=\linewidth]{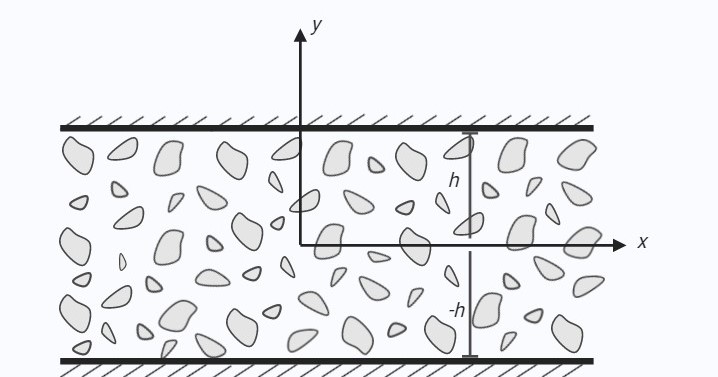}
\label{subfig:PM}
\caption{For porous medium 1D equation}
\end{subfigure}
\caption{Schematic representation of the domain}
\label{fig:schem}
\end{figure}
All of the selected models consider a unidirectional (1D) fully developed velocity field (velocity does not vary in the direction of flow) for fluid between parallel plane boundaries under the action of a pressure gradient parallel to the boundaries in the x-direction \cite{drazin_riley_2006}. The y-direction is perpendicular to the plane and separates the boundaries by $2h$.The center of the coordinate system is placed in the centerline of the domain $R = (-\infty,\infty) \times [-h,h] \in \mathbb{R}^2$ as shown in Fig.\ref{fig:schem}.

\subsection{1D Navier-Stokes equation and 1D porous medium model for a viscous flow}
The first 1D model of N-S equations describes a viscous flow or momentum diffusion, with constant density and viscosity without body forces. The dimensionless form of the equation is given by: 
\begin{equation}
\label{eq:DimNSx}
 \frac{d^2U}{d{\epsilon}^2}=-\frac{Re}{Ru}   
\end{equation}
where $U$ is the dimensionless velocity, $\epsilon$ the dimensionless space variable and $\frac{Re}{Ru}$ is a dimensionless comparison between internal forces and the diffusive process $\left(\frac{h^2\Delta P/L}{\mu_e}\right)$.The second order,linear, non-homogeneous ordinary differential equation (2nd,L,NH,ODE) eq.(\ref{eq:DimNSx}) describes the unidirectional flow of the diffusive process. The solution is obtained considering non-slip boundary conditions on both the plates $U(\epsilon=\pm 1)=0$:
 \begin{equation}
 \label{eq:DNSx}
     U=-\frac{Re}{2Ru}\left({{\epsilon}^2-1}\right)
\end{equation}

The Brinkman equation is a Darcian model considered a modification to Darcy's law that accounts for the viscous shear stresses that act on the fluid elements \cite{HAMDAN1994203}. This model has been used to analyze high-porosity porous media and has also been a subject of investigation, particularly about the boundary conditions at the solid and fluid- interface \cite{Durlofsky},\cite{Liu_2007}.
The term $\mu \mathbf{v}/k$ shows porous media's effect by the resistance to the flow exerted by the porosity of the medium through which the fluid passes (Darcy term). This term does not affect the non-linearity of governing partial differential equations \cite{changwoo}, the equation is given by:
\begin{equation}
\label{eq:GBri}
 0= -\nabla P + \mu_e {\nabla}^2 \mathbf{v}' - \frac{\mu}{k}\mathbf{v}' + \rho \mathbf{f}    
\end{equation}
where $\mathbf{v}'$ is the averaged velocity vector within the porous medium (filtration velocity vector), $\nabla p$ is the average pressure gradient $\mu_e$ is the effective viscosity of the fluid in the medium, $k$ is the permeability of the medium and $\mu$ is the viscosity of the fluid, the last term is considered the damping force caused by
the porous mass. Under the considerations for the simplified flow, eq.\ref{eq:GBri}) takes the form:
\begin{equation}
    \label{eq:Brink}
0= -\frac{dP}{dx}  + \mu_e \frac{d^2 v'_x}{dy^2}- \frac{\mu}{k} v'_x 
\end{equation}
where $v'_x$ is the x-component of the averaged velocity within the porous medium.
The dimensionless form of Eq.(\ref{eq:Brink}), considering $U$ as the dimensionless velocity and $\epsilon$ as dimensionless space variable, is given by:
\begin{equation}
\label{eq:DimenBri}
 0=- \frac{Re}{Ru}+\frac{d^2U}{d\epsilon^2}-\frac{\mu}{Da\mu_e}U 
\end{equation}
where $Da=\frac{h^2}{k}$ is the Darcy number, a ratio between the characteristic length of the domain and a macroscopic characteristic of the porous medium (permeability $k$). This number has been used to define flow regimes for the Brinkman equation, where the free flow is defined by $Da>1$, the porous flow with high permeability for $1>Da>10^ {-6}$ and porous flow or low permeability flow for $Da<10^ {-6}$ \cite{FEModel}.
Zaripov et al. \cite{Zaripov} recapped some mathematical relationships between $\frac{\mu}{\mu_e}$, mentioning that this ratio in porous media needs to be close to unity to satisfy the no-slip boundary conditions at the bounding walls. Based on it, $\frac{\mu}{\mu_e}=1$ is considered.
Eq.(\ref{eq:DimenBri}) is solved considering constant values for $Da$, $\mu=\mu_e$ and satisfying non-slip boundary conditions, the solution is given by:
\begin{equation}
\label{eq:BriSol}
U=-Da\frac{Re}{Ru}\left({\frac{1-e^{\frac{-2}{{\sqrt{Da}}}}}{e^{\frac{2}{\sqrt{Da}}}-e^{\frac{-2}{\sqrt{Da}}}} (e^{\frac{\epsilon+1}{\sqrt{Da}}}-e^{\frac{-\epsilon}{\sqrt{Da}}})+e^{-\frac{\epsilon +1}{\sqrt{Da}}}-1}\right) 
\end{equation}
\begin{figure}[htb!]
\centering
\includegraphics[width=0.65\linewidth]{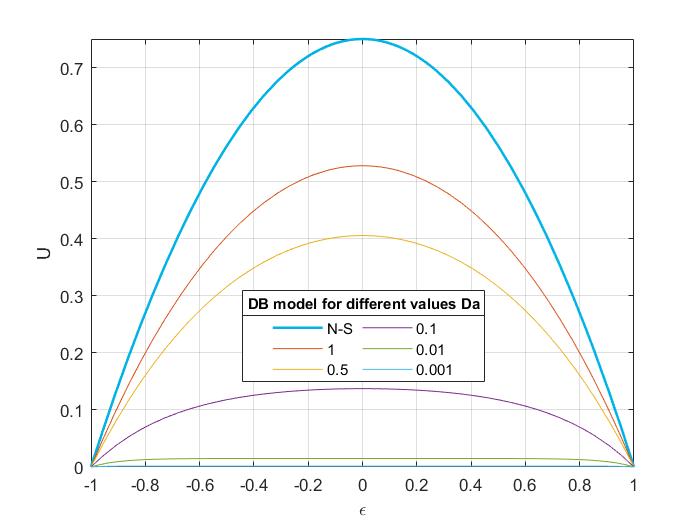}
\caption{Comparison between fluid flow through free space and a porous medium (Brinkman model)}
\label{fig:BriNS}
\end{figure}
The velocity profiles from eq.(\ref{eq:DNSx}) and eq.(\ref{eq:BriSol}) are plotted, for five different values of Darcy numbers $Da=\{1, 0.5, 0.1, 0.01, 0.001\}$, in Fig.(\ref{fig:BriNS}). The free-space model (N-S) presents higher velocity values than the porous medium model. The smaller the $Da$ number for the porous medium model, the slowest the flow.For $Da<1\times 10^{-2}$, the fluid barely flows compared to a free-space system, which means that the medium has a small permeability.The profiles agreed with Awartani et al. \cite{AWARTANI2005749} similar solution for a viscous dimensionless model and its comparison with Navier-Stokes equation, concluding that the effect of introducing a porous structure in the flow domain results in a reduction of the velocity relative to Navier-Stokes flow description.

\subsection{1D Navier-Stokes equations and 1D porous medium model for an inertial-viscous flow}
\label{Ss:ConvDiff}
The second model for the same system is a unidirectional diffusive-convective flow, known as the 1D convective-diffusion equation (inertial-viscous flow). The dimensionless component of the corresponding N-S equation is:
\begin{equation}
\label{eq:NSDCx}
\frac{d^2U}{d{\epsilon}^2}-Re\frac{dU}{d\epsilon}=-\frac{Re}{Ru}
\end{equation}
where $Re$ is the Reynolds number $\left(\frac{\rho h v_{max}}{\mu}\right)$, a comparison between inertial and viscous forces. Considering non-slip boundary conditions on both plates, the dimensionless analytical solution of the simplified model can be express as:
\begin{equation}
\label{eq:mc}
    U=-\frac{1}{Ru}\left[(\epsilon +1)-2Re\left(\frac{e^{Re (\epsilon)}-e^{-Re (\epsilon)}}{e^{Re}-e^{-Re}}\right)\right]
\end{equation}
where $Ru$ is the Ruark number, a comparison between convection and internal forces $(\frac{\rho V^2}{\Delta P})$.\\

The Darcy-Lapwood-Brinkman (DLB) equation is a non-Darcian model often used when viscous shear and macroscopic inertial effects are significant. This model is valid under macroscopic boundary and non-slip boundary conditions\cite{Merabet}. The equation includes the damping term $(\frac{mu}{k}\mathbf{v})$, also representing the Darcy resistance to motion and the convective term. The steady-state equation takes the form:
\begin{equation}
 \rho (\mathbf{v}' \cdot \nabla \mathbf{v}'   )= -\nabla P + \mu_e {\nabla}^2 \mathbf{v}' - \frac{\mu}{k}\mathbf{v}' + \rho \mathbf{f} 
\label{eq:DLBg}
\end{equation}
where $\mathbf{v}'$ is the averaged velocity vector, $\rho$ the constant density, $\mu_e$ the apparent viscosity that may depend on the geometry of the porous medium, $\mu$ the dynamic viscosity, $\nabla P$ the local pressure gradient, $k$ is the permeability of the porous medium and $\mathbf{f}$ the external force. Note that for an infinity value of permeability and $\mu_e = \mu$ DLB model is an averaged Navier-Stokes equations. For small values of permeability, Darcy's term is predominant $(Da<1\times 10^{-6})$. For 1D flow through a porous medium subjected to the domain presented in Fig.(\ref{fig:schem}\subref{subfig:PM}) and without external forces, Eq.(\ref{eq:DLBg}) takes the dimensionless form: 
\begin{equation}
\label{eq:DLBDimx}
 Re \frac{dU}{d\epsilon}= -\frac{Re}{Ru} + \frac{d^2U}{d\epsilon^2}- \frac{\mu}{\mu_e Da} U 
\end{equation}
\begin{figure}[htb!]
\centering
     \includegraphics[width=0.65\linewidth]{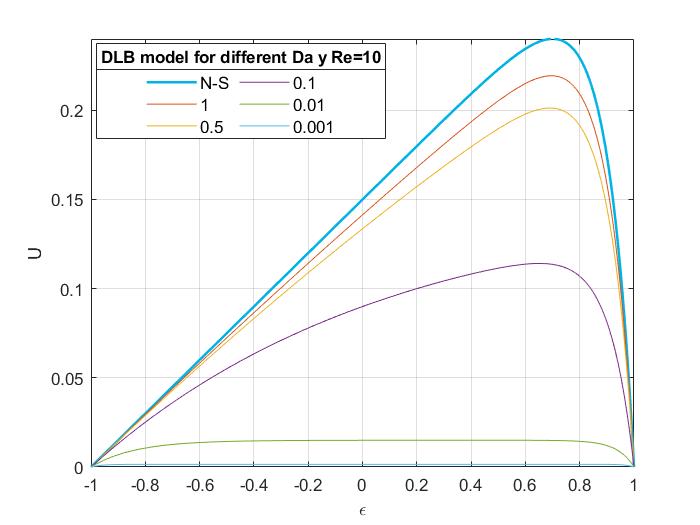}
    \caption{Convective-diffusive process}
    \label{fig:DLB-CD}
\end{figure}\\
Eq.(\ref{eq:DLBDimx}) is solved considering constant values for $\mu,\mu_e$ and $k$ and satisfying non-slip boundary conditions, the solution is given by:
\begin{equation}
\label{eq:DimenDLB-CD-Sol}
 U= {-\frac{Re}{Ru}Da}\left({\frac{(1-e^{-A})}{e^{B}-e^{A}}(e^{\frac{B(\epsilon+1)}{2}}-e^{\frac{A(\epsilon)}{2}})+e^{\frac{A(\epsilon+1)}{2}}-1}\right) 
\end{equation}\\
where $A=Re-\sqrt{(\frac{4}{Da})+Re^2}$, $B=Re+ \sqrt{(\frac{4}{Da})+Re^2}$.
The solutions Eq.(\ref{eq:DimenDLB-CD-Sol}) and Eq.(\ref{eq:mc}) are plotted in Fig.\ref{fig:DLB-CD}.The maximum value of the velocity is closest to the upper fixed plate due to the inertial effect. As the Darcy number decreases, the profile flattens, and the flow decreases. The smaller the Darcy number, the slowest the velocity of the fluid. 

\section{ Generalized Conformable 1D Navier-Stokes models}
\label{S:GCD_Model}
\begin{figure}[htb!]
\begin{subfigure}[b]{0.47\textwidth}
\centering
\includegraphics[width=\linewidth]{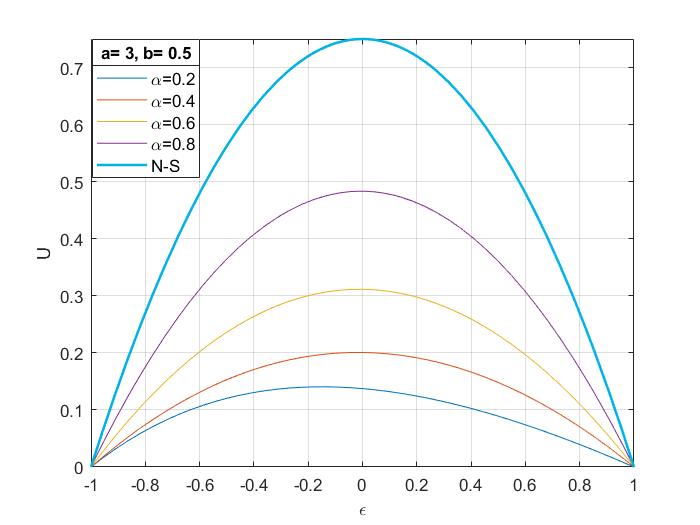}
\label{fig:Exp_a2.5}
\caption{a=3}
\end{subfigure}
\begin{subfigure}[b]{0.47\textwidth}
\centering
\includegraphics[width=\linewidth]{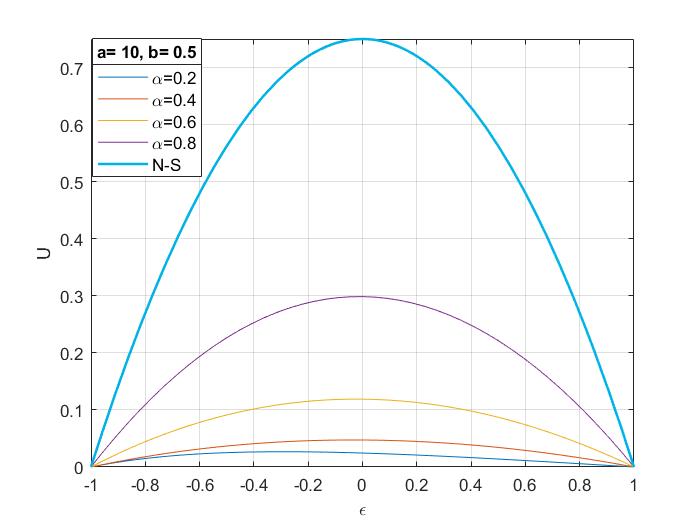}
\label{fig:E_a10}
\caption{a=10}
\end{subfigure}
\caption{Velocity profiles of diffusion conformable model for Poiseuille flow}
\label{fig:ExpFunct}
\end{figure}

The introduction of a space conformable derivative in a 1D Navier-Stokes equations are proposed to analyze the effect of this non-integer order derivative and its potential to reproduce some of the flows described in the previous section.\\
The initial analysis for the selection of the conformable function, and its impact on the description of the fluid flow, was made by an essential exponential function($ C (\alpha) e ^ {\alpha (1- \alpha)} $). This function did not accurately represent the flow, so a more generalized exponential function was proposed to suit better the flow description. The generalized function possesses two fitting parameters and the non-integer value $\alpha$, which allows more flexibility to describe the flow.  
The conformable fractional function proposed in this work, and based on the definitions given by Zhao et al.\cite{ZHAO}, is:
\begin{equation}
\label{eq:fcon}
 \psi\left({y},{\alpha}\right)={{a}}^{{(yb+1)}(1-{\alpha})}  
\end{equation}
where $a$, $b$ are the fitting parameters, $\alpha$ is the non-integer value of the derivative, and $y$ is the space variable in the differential equation. 
As proved in \cite{KHALIL14}, a conformable fractional derivative is directly related to the first derivative as:
\begin{equation}
\label{eq:CD1}
D^\alpha_\psi v_x (y)= {{a}}^{{(by+1)}(1-{\alpha})}\frac{dv_x}{dy}
\end{equation}
The same conformable function is applied to obtain second spacial derivative \cite{atangana15}:
\begin{equation}
\label{eq:2nD}
    D^\alpha_\psi (D^\alpha_\psi v_x (y))=a^{(yb+1)(1-\alpha)}\frac{d}{dy}\left(a^{(yb+1)(1-\alpha)}\frac{dv_x}{dy}\right)
\end{equation}
Replacing the first and second spatial derivatives in $x$ component of N-S (Eq.\ref{eq:DimNSx}), with  the dimensionless form of eq.(\ref{eq:CD1}) and eq.(\ref{eq:2nD}) respectively, the dimensionless spacial conformable fractional viscous model can be written as:
\begin{equation}
\label{eq:dconf2}   
\frac{d^2U}{d\epsilon^2}+bh(1-\alpha)\log(a)\frac{dU}{d\epsilon}=-\frac{Re}{Ru}a^{-2(1+b\epsilon h)(1-\alpha)}
\end{equation}
Analytical solution of Eq.(\ref{eq:dconf2}) considering non-slip boundary conditions can be express as:
\begin{equation}
\label{eq:DiffConf}
U=-\frac{1}{2A}\frac{Re}{Ru}\left(a^{-2\left(1-\alpha\right)\left(1+bh\epsilon\right)}-a^{-2(1-bh)(1-\alpha)}-C'_1(a^{-\left(1-\alpha\right)\left(1+bh\epsilon\right)}+a^{-\left(1-\alpha\right)\left(1-bh\right)})\right)
\end{equation}
where \begin{gather*}
A=\left((1-\alpha)bh\log(a)\right)^2 \\
C'_1=\frac{a^{-2(1+hb)\left(1-\alpha\right)}-a^{-2(1-hb)(1-\alpha)}}{a^{-(1+hb)\left(1-\alpha\right)}-a^{-(1-hb)(1-\alpha)}}
\end{gather*}
applying $\lim{\alpha\to 1}$, eq.(\ref{eq:DiffConf}) can represent the 1D Navier-Stokes equation classical solution. The velocity profiles of the conformable model are shown in Fig.(\ref{fig:ExpFunct}) for different values of fractional exponent $(\alpha=\{0.2, 0.4, 0.6,0.8\})$ and two different values of exponential base.Every conformable velocity profile is smaller than the classical N-S profile, and as the value of the base increases, the velocity decreases for a fixed value of $b$ parameter. 
\\

For the diffusive-convective model, the first and second spatial derivatives in Eq.(\ref{eq:NSDCx}) are replaced with the corresponding dimensionless fractional conformable spatial derivatives, obtaining:
\begin{equation}
\label{eq:cmconf}
\frac{dU}{d \epsilon^2}+\left[bh(1-\alpha)\log (a)-Re\right]\frac{dU}{d\epsilon}=-\frac{Re}{Ru}{a}^{-2(h\epsilon b+1)(1-\alpha)}
\end{equation}
Under non-slip boundary conditions on both plates, the analytical solution of Eq.(\ref{eq:cmconf}) takes the form: 
\begin{equation}
    \label{eq:SNSCD}
U=-\frac{1}{2B}\frac{Re}{Ru}\left(a^{-2\left(1-\alpha\right)\left(1+bh\epsilon\right)}-a^{-2(1-\alpha)(1-hb)}-C_1\frac{e^{-\beta \epsilon}-e^{\beta}}{e^{-\beta}-e^{\beta }}\right)
\end{equation}
where
\begin{gather*}
B=(1-\alpha)bh\log(a)\left((1-\alpha)bh\log(a)+Re\right)\\
\beta = (1-\alpha)bh\log(a)-Re\\ C_1={a^{-2\left(1-\alpha\right)\left(1+bh\right)}-a^{-2\left(1-\alpha\right)\left(1-hb\right)}}   
\end{gather*}
\begin{figure}[htb!]
\begin{subfigure}[b]{0.45\textwidth}
\centering
\includegraphics[width=\linewidth]{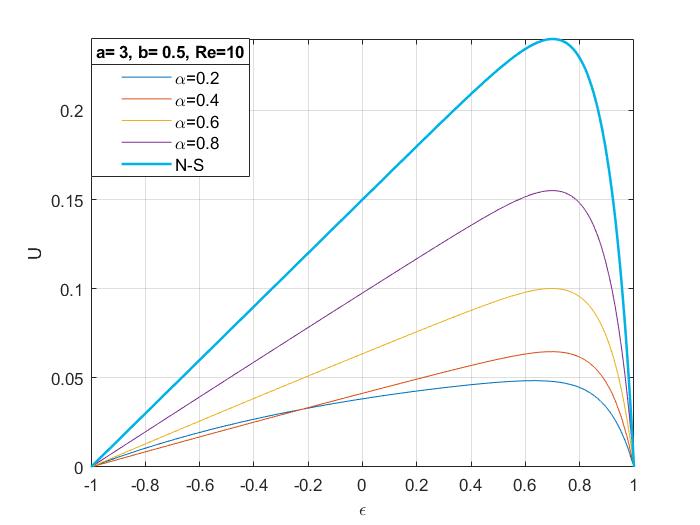}
\label{fig:CDRe5a2.5}
\caption{$Re=10$}
\end{subfigure}
\begin{subfigure}[b]{0.45\textwidth}
\centering
\includegraphics[width=\linewidth]{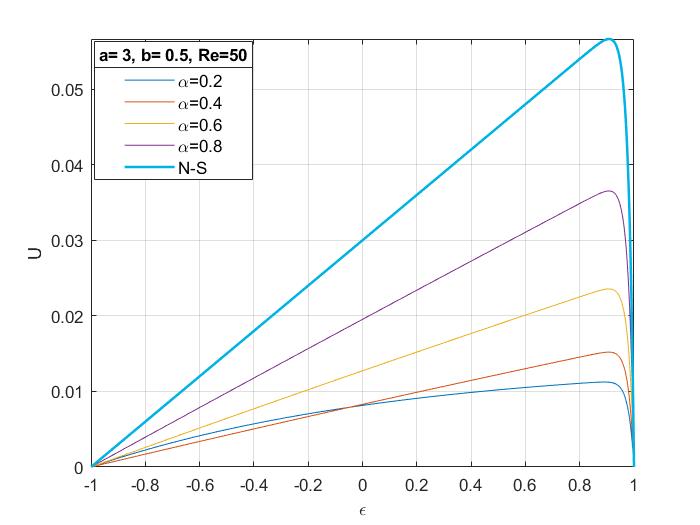}
\label{fig:CD_Re10}
\caption{$Re=50$}
\end{subfigure}
\caption{Velocity profiles of diffusion-convection conformable model for planar Poiseuielle flow}
\label{fig:ExpConDiff}
\end{figure}\\
Results displayed in Fig.\ref{fig:ExpConDiff} provide a comparison of velocity profiles between classical $(\alpha=1)$ and conformable models. Four values of the fractional exponent $(\alpha=\{0.3, 0.5, 0.7,0.8\})$ and two Reynolds numbers were selected to observe the effect of the inertial term on the velocity profile. The higher the $Re$, the slowest the flow, the convective term makes the flow slower.  In terms of conformable derivative order, the higher the order, the larger the velocity magnitude. As the convective term's value increases, the slowest the velocity for fixed values of the base of the exponential function $a$ and $b$.

Results on both conformable models are suitable to be compared to the corresponding porous medium, presenting reductions in the fluid's velocity. An optimization technique was applied to find the values of $a, b, \alpha$ for the conformable model representing Brinkman and DLB models, with a Matlab\textregistered \ minimum of constrained nonlinear multivariable function, it was possible to minimize the sum of the squared of residuals (SSR) between the porous medium model and the conformable model. The optimization was completed within the value of the software's optimal tolerance $(1\times 10^{-6})$.
\begin{figure}[htb!]
\begin{subfigure}[b]{0.47\linewidth}
\centering
\includegraphics[width=\textwidth]{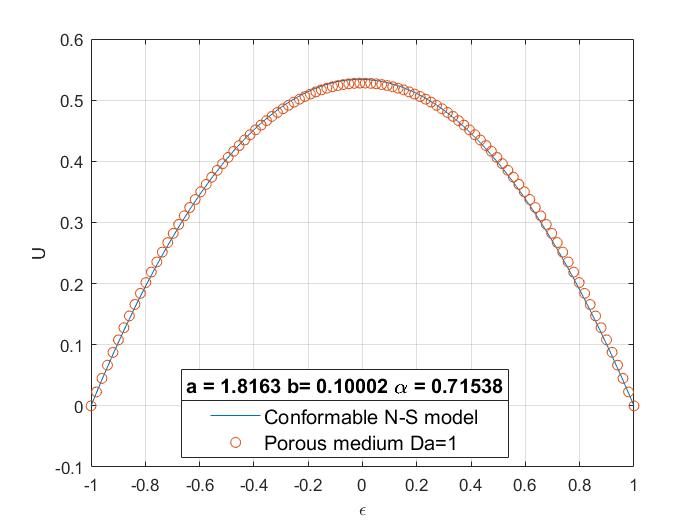}
\caption{$Da=1$}
\label{fig:BDa1}
\end{subfigure}
\begin{subfigure}[b]{0.47\textwidth}
\centering
\includegraphics[width=\textwidth]{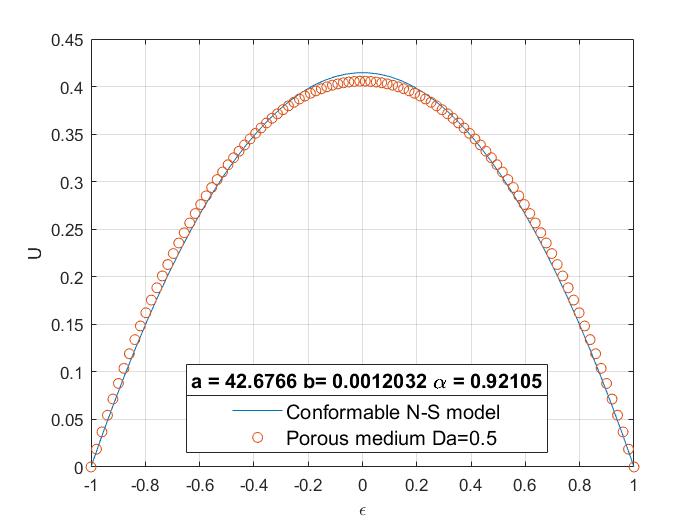}
\caption{$Da=0.5$}
\label{fig:BDa0.5}
\end{subfigure}
\begin{subfigure}[b]{0.47\textwidth}
\centering
\includegraphics[width=\textwidth]{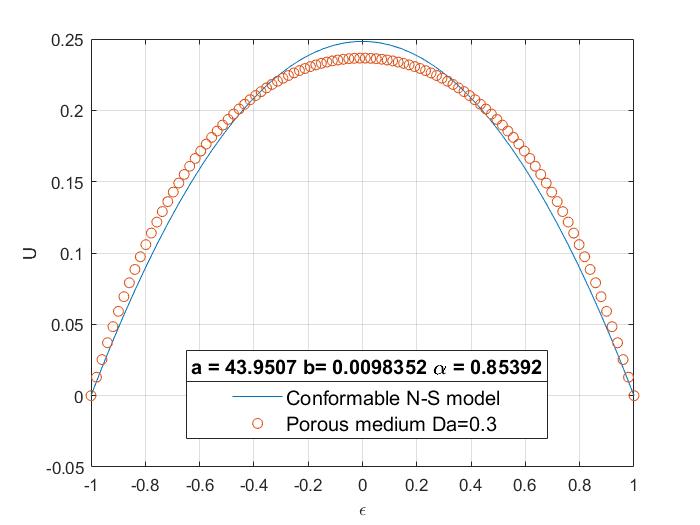}
\caption{$Da=0.3$}
\label{fig:BDa0.3}
\end{subfigure}\hfill
\begin{subfigure}[b]{0.47\textwidth}
\centering
\includegraphics[width=\textwidth]{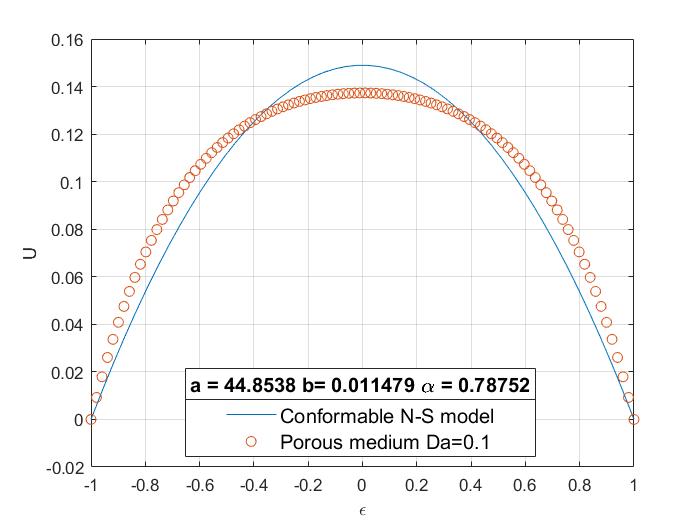}
\caption{$Da=0.01$}
\label{fig:BDa0.01}
\end{subfigure}
\caption{Conformable N-S diffusive model compared to Brinkman equation}
\label{FitD}
\end{figure}

\begin{figure}[htb!]
\begin{subfigure}[b]{0.45\linewidth}
\centering
\includegraphics[width=\linewidth]{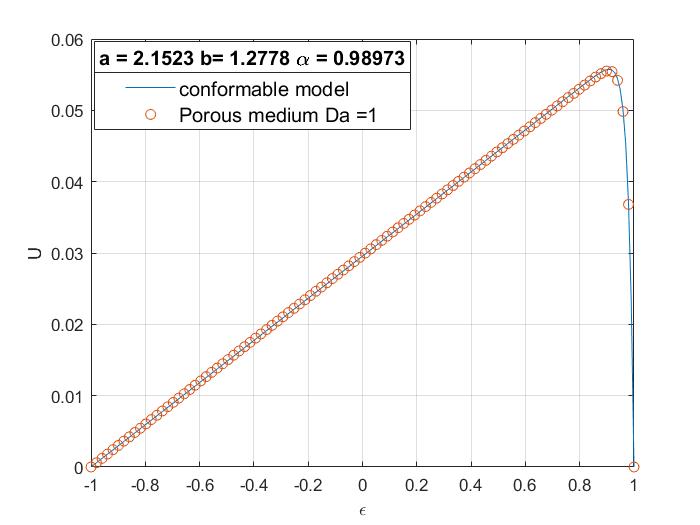}
\caption{$Da=1$}
\label{fig:DLRe50da1}
\end{subfigure}
\begin{subfigure}[b]{0.45\linewidth}
\centering
\includegraphics[width=\linewidth]{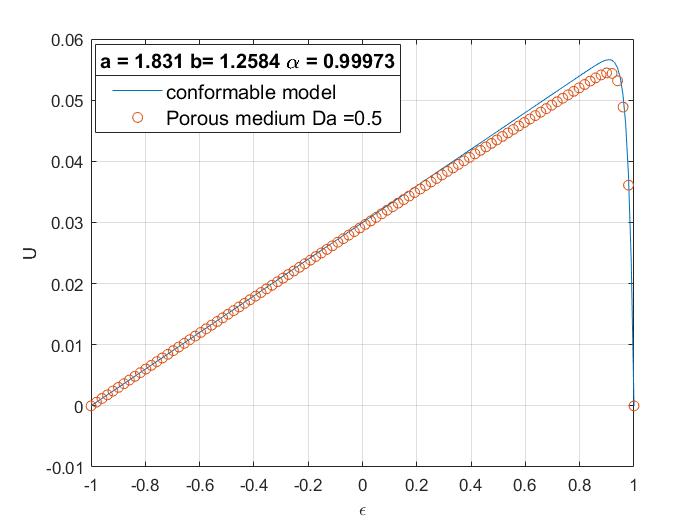}
\caption{$Da=0.5$}
\label{fig:DLBDa0.5Re50}
\end{subfigure}
\begin{subfigure}[b]{0.45\linewidth}
\centering
\includegraphics[width=\linewidth]{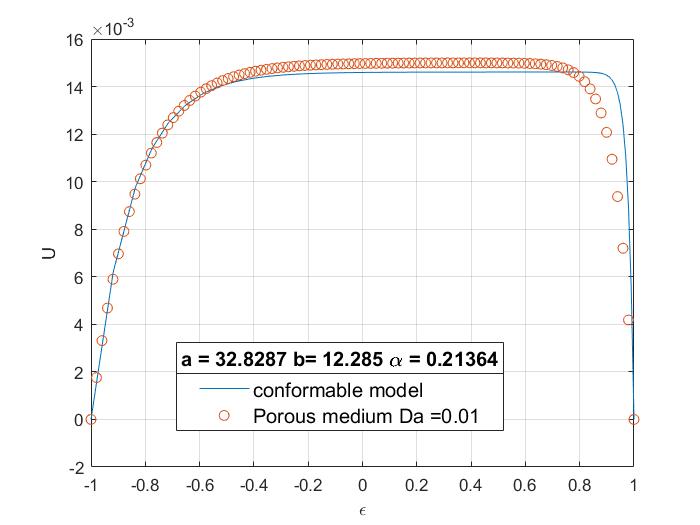}
\caption{$Da=0.01$}
\label{fig:DLBDa.01Re50}
\end{subfigure}\hfill
\begin{subfigure}[b]{0.45\linewidth}
\centering
\includegraphics[width=\linewidth]{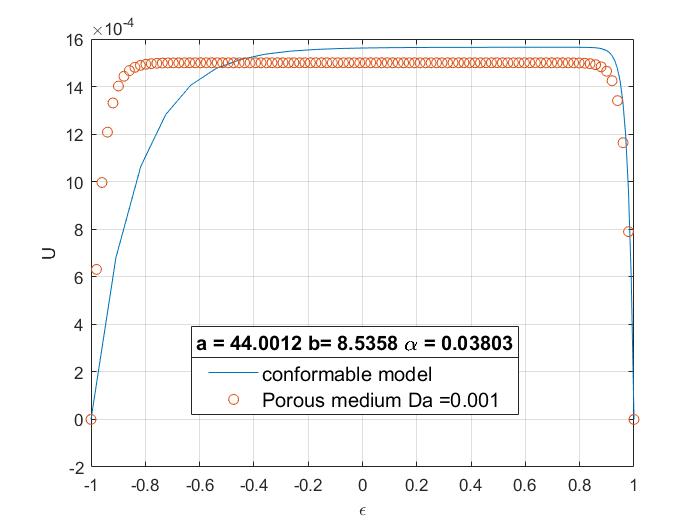}
\caption{$Da=0.001$}
\label{fig:DLBDa.001Re50}
\end{subfigure}
\caption{Conformable N-S diffusive model compared to DLB equation for $Re=50$ }
\label{fig:Re50}
\end{figure}

In Fig.\ref{FitD} the optimal parameters for the viscous conformable model are shown, and it is clear that the conformable model is ideally suited to be compared with the porous medium velocity profile. The value of fractional exponent $\alpha$ is less than one, which indicates that the conformable model is a better representation for this experimental laminar flow than the classical planar Poiseuille model. It is also showed that as the porous medium has a lower value of permeability, the conformable model with optimized generalized exponential function describes the process with less precision than the porous medium with high permeability values. The range of use of the conformable model as a porous medium model can be defined in terms of $0.1<Da\leq 1$.
\begin{figure}[htb!]
\begin{subfigure}[b]{0.45\linewidth}
\centering
\includegraphics[width=\linewidth]{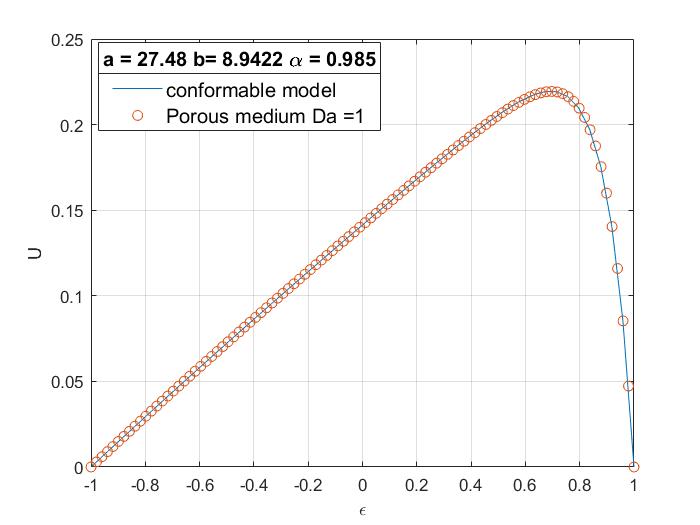}
\caption{$Da=1$}
\label{fig:DLBDa1}
\end{subfigure}
\begin{subfigure}[b]{0.45\linewidth}
\centering
\includegraphics[width=\linewidth]{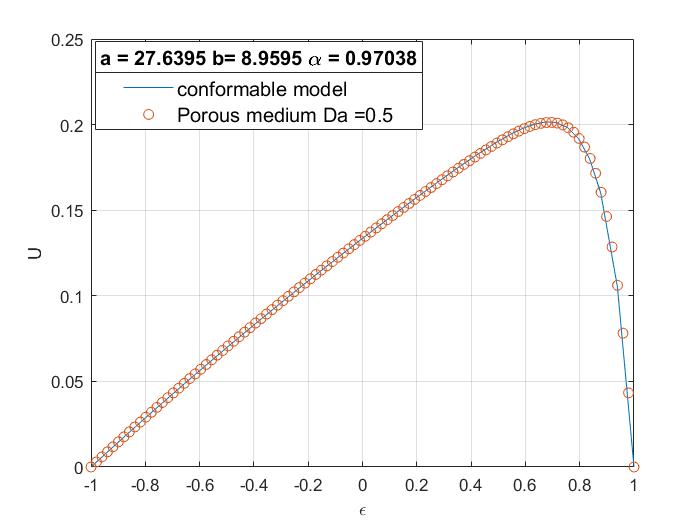}
\caption{$Da=0.5$}
\label{fig:DLBDa0.5}
\end{subfigure}
\begin{subfigure}[b]{0.45\linewidth}
\centering
\includegraphics[width=\linewidth]{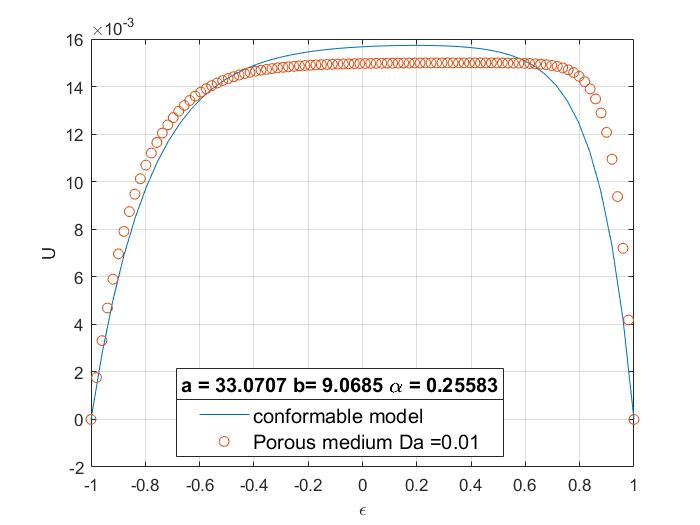}
\caption{$Da=0.01$}
\label{fig:DLBDa.01}
\end{subfigure}\hfill
\begin{subfigure}[b]{0.45\linewidth}
\centering
\includegraphics[width=\linewidth]{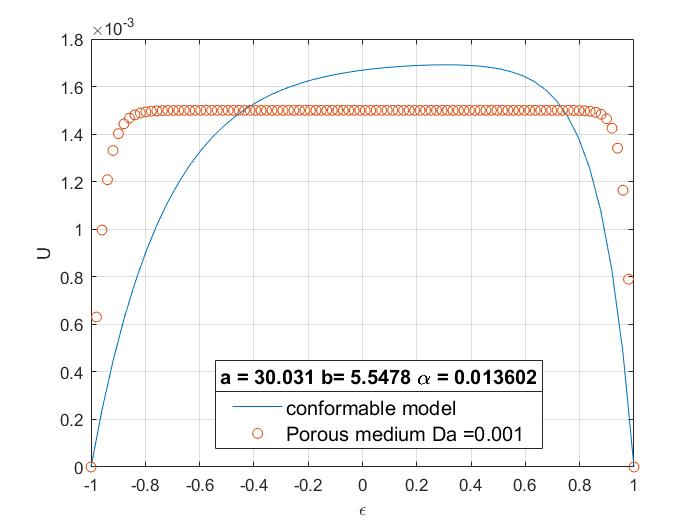}
\caption{$Da=0.001$}
\label{fig:DLBDa.001}
\end{subfigure}
\caption{Conformable N-S diffusive model compared to DLB equation for $Re=10$ }
\label{fig:Re10}
\end{figure}
In Fig.\ref{fig:Re10} and Fig.\ref{fig:Re50} the conformable model is compared to DLB model, including the optimization of parameters, for two Reynolds numbers ($Re=\{10,50\}$). As the Darcy number decreases ($Da < 0.01$), the flow becomes more uniform across the section, almost like a plug flow.  For higher values of the Reynolds number, the slowest the flow. The minimum value of velocity in the cases presented is shown in Fig.\ref{fig:DLBDa.001Re50}, in which the conformable model presents significant deviations for DLB. The range of validity of the conformable model for an inertial-viscous flow, for both $Re$ numbers, is for $0.01<Da\leq 1$.\\
Through the analysis of the Darcy number, the range of applicability of the conformable model was obtained. For $Da>10^{-2}$ the diffusive conformable model can represent a Darcian flow. This value indicates that the permeability of the porous medium has to be a hundred times smaller than the characteristic length of the space through which the flow takes place. For the process with inertial effects, the value of Darcy number can be even smaller $Da>10^{-3}$.

The introduction of the space conformable derivative does not affect the simplified N-S models' linearity, and traditional differential equations solving methods can be applied to conformable models. The analytical solutions are not significantly more complicated than the classical porous media solutions. Additionally, the dimensionless form of the equations did not require particular fractional or conformable dimensionless definitions. These features place the conformable models as a new approach to model flow through heterogeneous spaces.
The optimized generalized exponential function selected can describe both Darcian and non-Darcian flows. Additionally, the conformable model reproduced the flow through the porous medium without considering the empirical term found in Brinkman and DLB models (to describe the resistance to the flow exerted by the porosity of the medium), suggesting that flows through high porosity mediums can be expressed without any specific macroscopic data of the porous medium or an averaged velocity component.  Consequently, the possibility of describing simple flows through a porous medium can be more accessible with a conformable spatial model, in which discussions about the characteristics of the medium $(k, \frac{\mu}{\mu_e})$ are not needed. 
With the results obtained in this work, it is possible to connect the conformable spatial derivative to a geometrical and physical interpretation, showing the conformable derivative's potential to transform a free space medium into an inhomogeneous one, giving a new approach for porous medium fluid flow.    

\section{Conclusions}
Analytical dimensionless solutions for the 1D Brinkman model and the 1D Darcy-Lapwood-Brinkman model, describing the flow in a porous medium between two fixed walls, were compared to a 1D spatial conformable N-S models. A generalized exponential function with three parameters $(a,b,\alpha)$ was proposed to compare with the porous medium models. The optimized conformable exponential function with $\alpha<1$ successfully represented both Darcian and non-Darcian flows, within a range of Darcy numbers $(1\times10^{-2}<Da\leq 1)$, in which the validity of both models has been proved. The obtained results showed that the generalized conformable model could describe the flow through a porous medium without including a Darcy term or macroscopical porous characteristics. 
Depending on the selected generalized exponential function, the conformable spatial model can represent a specific new geometrical space and different physical interpretations. The conformable model can transform a free space medium into an inhomogeneous one and provide a new approach for porous medium flow descriptions.  
\label{S:Concl}

\section*{Acknowledgements}
This study was supported by the Mexican National Council for Science and Technology (CONACyT) (Grant: 325202), Universidad Iberoamericana, DNVP, and FICSAC.  

\bibliographystyle{abbrv}
\bibliography{exportlist.bib}

\begin{thebibliography}{10}

\bibitem{RN207}
T.~Abdeljawad.
\newblock On conformable fractional calculus.
\newblock {\em Journal of Computational and Applied Mathematics}, 279:57--66,
  2015.

\bibitem{AKBULUT2018876}
A.~Akbulut and M.~Kaplan.
\newblock Auxiliary equation method for time-fractional differential equations
  with conformable derivative.
\newblock {\em Computers and Mathematics with Applications}, 75(3):876 -- 882,
  2018.

\bibitem{ALHARBI201918}
F.~M. Alharbi, D.~Baleanu, and A.~Ebaid.
\newblock Physical properties of the projectile motion using the conformable
  derivative.
\newblock {\em Chinese Journal of Physics}, 58:18 -- 28, 2019.

\bibitem{anderson2015newly}
D.~R. Anderson and D.~J. Ulness.
\newblock Newly defined conformable derivatives.
\newblock {\em Adv. Dyn. Syst. Appl}, 10(2):109--137, 2015.

\bibitem{AtanganaBook}
A.~Atangana.
\newblock {\em Derivative with a New Parameter: Theory, Methods and
  Applications}.
\newblock Academic Press, 10 2015.

\bibitem{atangana15}
A.~Atangana, D.~Baleanu, and A.~Alsaedi.
\newblock New properties of conformable derivative.
\newblock {\em Open Mathematics}, 1(open-issue), 2015.

\bibitem{benkhettou2016conformable}
N.~Benkhettou, S.~Hassani, and D.~F. Torres.
\newblock A conformable fractional calculus on arbitrary time scales.
\newblock {\em Journal of King Saud University-Science}, 28(1):93--98, 2016.

\bibitem{CHAUDHARY20}
M.~Chaudhary, R.~Kumar, and M.~K. Singh.
\newblock Fractional convection-dispersion equation with conformable derivative
  approach.
\newblock {\em Chaos, Solitons and Fractals}, 141:110426, 2020.

\bibitem{Chen}
J.-S. Chen and C.-W. Liu.
\newblock Generalized analytical solution for advection-dispersion equation in
  finite spatial domain with arbitrary time-dependent inlet boundary condition.
\newblock {\em Hydrology and Earth System Sciences}, 15:2471--2479, 08 2011.

\bibitem{CLARKE20}
D.~A. Clarke, F.~Dolamore, C.~J. Fee, P.~Galvosas, and D.~J. Holland.
\newblock Investigation of flow through triply periodic minimal
  surface-structured porous media using {MRI} and {CFD}.
\newblock {\em Chemical Engineering Science}, 231:116264, 2021.

\bibitem{drazin_riley_2006}
P.~G. Drazin and N.~Riley.
\newblock {\em The {Navier-Stokes Equations}: A Classification of Flows and
  Exact Solutions}.
\newblock London Mathematical Society Lecture Note Series. Cambridge University
  Press, 2006.

\bibitem{Durlofsky}
L.~Durlofsky and J.~F. Brady.
\newblock Analysis of the {Brinkman} equation as a model for flow in porous
  media.
\newblock {\em The Physics of Fluids}, 30(11):3329--3341, 1987.

\bibitem{HAMDAN1994203}
M.~Hamdan.
\newblock Single-phase flow through porous channels a review of flow models and
  channel entry conditions.
\newblock {\em Applied Mathematics and Computation}, 62(2):203 -- 222, 1994.

\bibitem{HADMAN91}
M.~H. Hamdan and R.~M. Barron.
\newblock Analysis of the {Darcy-Lapwood} and the {Darcy-Lapwood-Brinkman}
  models: significance of the laplacian.
\newblock {\em Applied Mathematics and Computation}, 44(2):121--141, 1991.

\bibitem{Merabet}
N.~M. H. S.~M. Hamdan.
\newblock Analytical approach to the {Darcy–Lapwood–Brinkman} equation.
\newblock {\em Applied Mathematics and Computation}, 196, 2008.

\bibitem{hosseini2017new}
K.~Hosseini, A.~Bekir, M.~Kaplan, and {\"O}.~G{\"u}ner.
\newblock On a new technique for solving the nonlinear conformable
  time-fractional differential equations.
\newblock {\em Optical and Quantum Electronics}, 49(11):343, 2017.

\bibitem{hyder2020exact}
A.-A. Hyder and A.~H. Soliman.
\newblock Exact solutions of space-time local fractal nonlinear evolution
  equations: A generalized conformable derivative approach.
\newblock {\em Results in Physics}, page 103135, 2020.

\bibitem{IYIOLA2017}
O.~Iyiola, O.~Tasbozan, A.~Kurt, and Y.~Çenesiz.
\newblock On the analytical solutions of the system of conformable
  time-fractional {Robertson} equations with 1-d diffusion.
\newblock {\em Chaos, Solitons and Fractals}, 94:1 -- 7, 2017.

\bibitem{KANAUN2020}
S.~Kanaun and V.~Levin.
\newblock Fluid filtration through the media with random sets of crack-like
  inclusions.
\newblock {\em International Journal of Engineering Science}, 156:103370, 2020.

\bibitem{changwoo}
C.~Kang and P.~Mirbod.
\newblock Porosity effects in laminar fluid flow near permeable surfaces.
\newblock {\em Phys. Rev. E}, 100:013109, Jul 2019.

\bibitem{KHALIL14}
R.~Khalil, M.~{Al Horani}, A.~Yousef, and M.~Sababheh.
\newblock A new definition of fractional derivative.
\newblock {\em Journal of Computational and Applied Mathematics}, 264:65 -- 70,
  2014.

\bibitem{khoei2021}
A.~Khoei and S.~Saeedmonir.
\newblock Computational homogenization of fully coupled multiphase flow in
  deformable porous media.
\newblock {\em Computer Methods in Applied Mechanics and Engineering},
  376:113660, 2021.

\bibitem{Lei}
G.~Lei, N.~Cao, D.~Liu, and H.~Wang.
\newblock A non-linear flow model for porous media based on conformable
  derivative approach.
\newblock {\em Energies}, 11(11), 2018.

\bibitem{li2020existence}
S.~Li, S.~Zhang, and R.~Liu.
\newblock The existence of solution of diffusion equation with the general
  conformable derivative.
\newblock {\em Journal of Function Spaces}, 2020, 2020.

\bibitem{Liu_2007}
H.~Liu, P.~Patil, and U.~Narusawa.
\newblock On {Darcy-Brinkman Equation}: {Viscous Flow Between Two Parallel
  Plates Packed with Regular Square Arrays of Cylinders}.
\newblock {\em Entropy}, 9(3):118–131, Sep 2007.

\bibitem{maruvsic20}
E.~Maru{\v{s}}i{\'c}-Paloka and I.~Pa{\v{z}}anin.
\newblock Homogenization and singular perturbation in porous media.
\newblock {\em Communications on Pure \& Applied Analysis}, page~1, 2020.

\bibitem{AWARTANI2005749}
M.Awartani and M.~Hamdan.
\newblock Fully developed flow through a porous channel bounded by flat plates.
\newblock {\em Applied Mathematics and Computation}, 169(2):749 -- 757, 2005.

\bibitem{Neale}
G.~Neale and W.~Nader.
\newblock Practical significance of {Brinkman's} extension of {Darcy's} law:
  Coupled parallel flows within a channel and a bounding porous medium.
\newblock {\em The Canadian Journal of Chemical Engineering}, 52(4):475--478,
  1974.

\bibitem{FEModel}
M.~Parvazinia, V.~Nassehi, R.~Wakeman, and M.~H.~R. Ghoreishy.
\newblock Finite element modelling of flow through a porous medium between two
  parallel plates using the brinkman equation.
\newblock {\em Transport in Porous Media}, 63:71--90, 01 2006.

\bibitem{NSCtim}
J.~Shao, B.~L. Guo, and L.~L. Duan.
\newblock Analytical study of the two-dimensional time-fractional
  {Navier-Stokes} equations.
\newblock {\em Journal of Applied Analysis and Computation}, 9(5):1999--2022,
  2019.

\bibitem{sherwood20}
J.~D. Sherwood.
\newblock Unsteady flow adjacent to an oscillating or impulsively started
  porous wall.
\newblock {\em Journal of Fluid Mechanics}, 894:A1, 2020.

\bibitem{Slattery}
J.~C. Slattery.
\newblock Single-phase flow through porous media.
\newblock {\em AIChE Journal}, 15(6):866--872, 1969.

\bibitem{WHITAKER1966291}
S.~Whitaker.
\newblock The equations of motion in porous media.
\newblock {\em Chemical Engineering Science}, 21(3):291 -- 300, 1966.

\bibitem{YANG2020106330}
S.~Yang, X.~Chen, L.~Ou, Y.~Cao, and H.~Zhou.
\newblock Analytical solutions of conformable advection–diffusion equation
  for contaminant migration with isothermal adsorption.
\newblock {\em Applied Mathematics Letters}, 105:106330, 2020.

\bibitem{yang2018}
S.~Yang, L.~Wang, and S.~Zhang.
\newblock Conformable derivative: Application to non-darcian flow in
  low-permeability porous media.
\newblock {\em Applied Mathematics Letters}, 79:105--110, 2018.

\bibitem{Yang2019AnalyticalSO}
S.~Yang, H.~Zhou, S.~Zhang, and L.~Wang.
\newblock Analytical solutions of advective-dispersive transport in porous
  media involving conformable derivative.
\newblock {\em Appl. Math. Lett.}, 92:85--92, 2019.

\bibitem{yavuz2018conformable}
M.~Yavuz and B.~Ya{\c{s}}k{\i}ran.
\newblock Conformable derivative operator in modelling neuronal dynamics.
\newblock {\em Applications \& Applied Mathematics}, 13(2), 2018.

\bibitem{yepez2018first}
H.~Y{\'e}pez-Mart{\'\i}nez, J.~G{\'o}mez-Aguilar, and A.~Atangana.
\newblock First integral method for non-linear differential equations with
  conformable derivative.
\newblock {\em Mathematical Modelling of Natural Phenomena}, 13(1):14, 2018.

\bibitem{Zaripov}
S.~Zaripov, R.~Mardanov, and V.~Sharafutdinov.
\newblock Determination of {Brinkman Model Parameters Using Stokes Flow Model}.
\newblock {\em Transport in Porous Media}, 130, 11 2019.

\bibitem{ZHAO}
D.~Zhao, X.~Pan, and M.~Luo.
\newblock A new framework for multivariate general conformable fractional
  calculus and potential applications.
\newblock {\em Physica A}, 5(10):271--280, 2018.

\bibitem{AnomDiffConfor}
H.~Zhou, S.~Yang, and S.~Zhang.
\newblock Conformable derivative approach to anomalous diffusion.
\newblock {\em Physica A: Statistical Mechanics and its Applications},
  491:1001--1013, 02 2018.

\bibitem{Burgersfluid}
Y.~Çenesiz, D.~Baleanu, A.~Kurt, and O.~Tasbozan.
\newblock New exact solutions of burgers’ type equations with conformable
  derivative.
\newblock {\em Waves in Random and Complex Media}, 27(1):103--116, 2017.

\end{thebibliography}







\end{document}